\documentclass[aps,prl,twocolumn,superscriptaddress,showpacs]{revtex4-1}
\usepackage{bm,graphicx,amsmath,amssymb,color,placeins,tikz,pgflibraryarrows,braket,colortbl,xcolor,mathtools}
\def\k{\kappa}

\def\bG{{\bf G}}
\def\bq{{\bf q} }
\def\bQ{{\bf Q} }
\def\bR{{\bf R} }
\def\bt{{\bm \tau}}

\def\<{\langle}
\def\>{\rangle}

\let\hide\iffalse
\usepackage[caption=false]{subfig}

\begin{document}

\title{Efficient first-principles methodology for the calculation \\ of the all-phonon inelastic scattering in solids}
\author{Marios Zacharias}
\email{marios.zacharias@cut.ac.cy}
\affiliation{Department of Mechanical and Materials Science
  Engineering, Cyprus University of Technology, P.O. Box 50329, 3603
  Limassol, Cyprus}
\author{H\'el\`ene Seiler}
\affiliation{Fritz-Haber-Institut, Physical Chemistry Department, Berlin, 14195, Germany}
\author{Fabio Caruso}
\affiliation{Institut f\"ur Theoretische Physik und Astrophysik, Christian-Albrechts-Universit\"at zu Kiel, D-24098 Kiel, Germany}
\author{Daniela Zahn}
\affiliation{Fritz-Haber-Institut, Physical Chemistry Department, Berlin, 14195, Germany}
\author{Feliciano Giustino}
\affiliation{ Oden Institute for Computational Engineering and Sciences, The University of Texas at Austin,
Austin, Texas 78712, USA%\\This line break forced with \textbackslash\textbackslash
}%
\affiliation{Department of Physics, The University of Texas at Austin, Austin, Texas 78712, USA}
\author{Pantelis C. Kelires}
\affiliation{Department of Mechanical and Materials Science
  Engineering, Cyprus University of Technology, P.O. Box 50329, 3603
  Limassol, Cyprus}
\author{Ralph Ernstorfer}
\email{ernstorfer@fhi-berlin.mpg.de}
\affiliation{Fritz-Haber-Institut, Physical Chemistry Department, Berlin, 14195, Germany}
\date{\today}

\begin{abstract}
Inelastic scattering experiments are key methods for mapping the full dispersion of 
fundamental excitations of solids in the ground as well as non-equilibrium states. 
A quantitative analysis of inelastic scattering in terms of phonon excitations 
requires identifying the role of multi-phonon processes. 
Here, we develop an efficient first-principles methodology for calculating 
the {\it all-phonon} quantum mechanical structure factor of solids. We demonstrate our method 
by obtaining excellent agreement between measurements and calculations of the diffuse 
 scattering patterns of black phosphorus, showing that multi-phonon 
 processes play a substantial role. The present approach constitutes a step towards the interpretation
of static and time-resolved electron, X-ray, and neutron inelastic scattering data. 
\end{abstract}

\maketitle

Inelastic scattering experiments in solids have a long history and have been the subject
of intensive research for almost a century. Originally employed to understand 
atomic vibrations~\cite{Born_1942}, this type of experiments reveals the full dispersion relations
of fundamental collective excitations like 
phonons~\cite{Nilson_1971, Holt1999, Muller_2001, Shukla_2003, Baron_2004, Xu2005, Matteo_2007, Reznik_2009, LeTacon2013}, 
plasmons~\cite{Abajo_2010} and spins~\cite{Moon_1969, Bowman_2019}, as well as of 
localized excitations like polarons~\cite{Adams_2000,Pengcheng_2000}, or excitons~\cite{Jinhua_2020}.
Since the development of time-resolved diffraction~\cite{Gerard_1982,Lunney_1986},
standard techniques of inelastic scattering have been gradually taken to the ultrafast 
time-domain~\cite{Lindenberg_2000, Cavalleri_2000, Trigo2010, Trigo2013,Harb_2016,
Waldecker_2016, Waldecker2017, Nicholson2018, Wall2018, Stern_2018, Konstantinova_2018, Teitelbaum_2018,
 Na2019, Zahn_2020, Seiler2021}. In this regime, new scattering signatures 
emerge, reflecting intriguing nonequilibrium physics that arise from many-body interactions~\cite{Caruso_2021}. 
This wealth of information is obtained by analyzing the inelastic contribution 
to the total scattering signal. However, scattering patterns are usually dominated
by inelastic interactions with phonons, for which a full description, beyond the standard one-phonon structure 
factor~\cite{Grosso_Pastori_book}, is critical to (i) improve the analysis of phonon excitations and 
(ii) extract other excitation signals with small cross section. 

\begin{figure}[b!]
\includegraphics[width=0.40\textwidth]{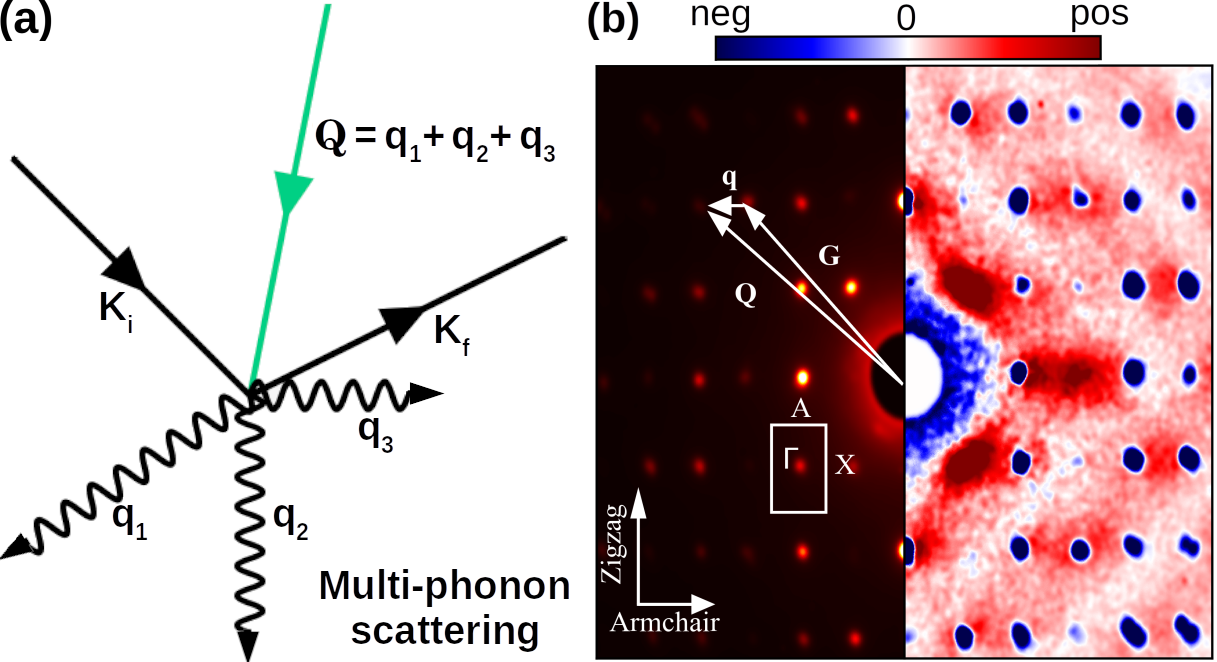}
\caption{
(a) Schematic of a multi-phonon scattering process. ${\bf K}_i$ and ${\bf K}_f$ denote the 
beam wavevectors before and after inelastic scattering. The momentum transfer to 
the crystal is $\hbar \bQ = \hbar( \, \bq_1 + \bq_2 + \bq_3)$, 
where $\bQ$ and $\bq$ are the scattering and reduced phonon wavevectors.
(b) Left: raw scattering pattern of bP as collected 
on the detector. $\bG$ represents a Bragg peak vector.  
The Brillouin zone with the $\Gamma$, A, and X  high-symmetry points together with the armchair and zigzag directions 
are indicated. Right: difference scattering pattern, $\Delta I(\bQ, t= 50 {\rm \, ps} )$, described in the
main text. 
}
\label{fig1}
\end{figure}

The quantum theory describing inelastic scattering from phonons, also known as thermal 
diffuse scattering, has been developed by Laval~\cite{Laval_1939}, Born~\cite{Born_1942}, and 
James~\cite{James_1948} (LBJ) and is included in several solid-state
physics textbooks~\cite{Maradudin_Weiss,Grosso_Pastori_book,Zhong_Lin_Wang,FG_Book}.
The appealing characteristic of the LBJ theory is that
one- and multi-phonon processes are treated on the same footing, allowing for 
the evaluation of the all-phonon scattering intensity using a single compact expression. 
Albeit this theory and the principle of multi-phonon scattering [c.f. Fig. 1(a)] are well 
established for decades~\cite{Sjolander_1958,Dawidowski_1998,Baron_2007,Kuroiwa_2008,Baron2014,Wehinger_2017}, 
current {\it ab-initio} calculations of diffuse diffraction~\cite{Konstantinova_2018,Krishnamoorthy_2019,Cotret_2019,
Bowman_2019, Bernardi_2021,Maldonado_2020,Seiler2021,Otto2021} only account for 
the one-phonon structure factor. 
This approach becomes problematic whenever scattering
wavevectors in high order Brillouin zones and/or high temperatures are
of interest. In such cases, multi-phonon effects hinder an accurate
analysis of experimental data, e.g., for extracting phonon dispersions
and nonequilibrium phonon populations.
%This practise is typically problematic whenever scattering wavevectors in high order Brillouin zones
%and/or high temperatures are of interest, hindering an accurate analysis of experimental data, 
%e.g., for extracting phonon dispersions and nonequilibrium phonon populations.

In this Letter, we develop an efficient first-principles method 
for the calculation of the one-, multi-, and all-phonon scattering in solids, 
relying on the LBJ formalism. We demonstrate the predictive power of this 
methodology by performing electron diffraction measurements and
calculations of the scattering intensity of black phosphorus (bP). 
Theory and measurements are in striking agreement, confirming the decisive role of 
multi-phonon interactions in reproducing experiment for a large range of scattering wavevectors. 
We also evaluate the fraction of the thermal energy transfer due to multi-phonon 
excitations and find up to 30\% contribution in the temperature range 100 -- 500~K.
The computational approach developed here carries general validity and 
can be employed to analyze electron, X-ray, and neutron inelastic scattering of any material, as long as
the kinematic approximation holds.  

To measure the thermal diffraction signals of bP we performed Femtosecond Electron Diffuse
Scattering (FEDS)~\cite{Waldecker2017, Stern_2018, Cotret_2019, Seiler2021} 
and focus on the hot, thermalized phonon populations resulting from photo-excitation. 
The experimental setup is described in the parallel paper, Ref.~\cite{Zacharias_arxiv_2021}.
In Fig.~\ref{fig1}(b) we show a typical thermal difference pattern of
bP obtained as $\Delta I(\bQ, t) = I(\bQ, t) - I(\bQ, t')$, 
where $I(\bQ, t')$ and $I(\bQ,t)$ are the average scattering intensities prior to photoexcitation 
measured at time $t'$  and at a pump-probe delay $t = 50$~ps, at which the bP lattice 
has acquired a quasi-thermalized state~\cite{Zahn_2020}.  
Each scattering wavevector $\bQ$ on this pattern can be generally expressed as 
$\bQ = \bG \pm \sum_{i} \bq_i$, 
where $\bG$ is a Bragg vector and $\bq_i$ are the reduced wavevectors of the phonons
involved in the absorption or emission processes [see Fig.~\ref{fig1}(a)]. 
The negative and positive $\Delta I(\bQ)$ are marked in blue and red, respectively. 
Bragg (elastic) scattering intensity appears as negative owing to the Debye-Waller damping, 
while diffuse (inelastic) scattering appears as positive due to larger phonon 
populations at thermalization~\cite{Zacharias_arxiv_2021}. 

In the framework of the adiabatic LBJ theory, the scattering intensity arising from an instantaneous 
atomic configuration is given by the amplitude of the total scattering factor, as a consequence of 
the kinematic limit (or first Born approximation)~\cite{Van_Hove_1954, Zhong_Lin_Wang}. Formally, 
this approximation involves truncating the Born expansion of the Lippmann–Schwinger 
equation~\cite{Lippman_1950} up to the first order in the interaction potential. Physically, this 
is valid for weak probe-crystal interactions and neglects multiple scattering, 
 i.e. the beam undergoes a single interaction event. Under these conditions,
the energy attenuation of the incident beam is due to inelastic scattering from 
lattice vibrations and the temperature dependence of the collected intensity 
can be evaluated as a canonical ensemble average with the electrons in their ground state.

Employing the harmonic approximation, the all-phonon LBJ scattering intensity at temperature $T$ 
can be calculated from the following compact expression~\cite{Zacharias_arxiv_2021,Xu2005}: 
\begin{eqnarray}\label{eqa1.7}
I_{\rm all}(\bQ,T) &=& N_p  \sum_{p} \sum_{\k \k'} f_\k (\bQ) f^*_{\k'} (\bQ)
    e^{ i \bQ \cdot [\bR_p + \bt_\k - \bt_{\k'} ] } \nonumber 
    \\ &\times&  e^{-W_{\k} (\bQ,T)} \, e^{-W_{\k'} (\bQ,T)} \, e^{ P_{p,\k\k'} (\bQ,T)},
\end{eqnarray}
where $\bt_\k$ represents the equilibrium positions of atom $\k$ in unit cell $p$, 
and $\bR_p$ the position vectors of $N_p$ unit cells contained in a periodic supercell. 
$f_\k (\bQ)$ denotes the atomic scattering amplitude, $W_{\k} (\bQ,T)$ is the exponent 
of the Debye-Waller factor~\cite{Zacharias_arxiv_2021}, 
and $P_{p,\k\k'} (\bQ,T)$ is the exponent of the phononic factor given by~\cite{Zacharias_arxiv_2021}:
\begin{eqnarray}\label{eqa1.8_b}
&& P_{p,\k\k'} (\bQ,T)  = \\ && \frac{M_0 N^{-1}_p}{\sqrt{M_\k M_{\k'}}} \sum_{\bq \nu }   u^2_{\bq \nu} 
 \text{Re}\Big[ \bQ \cdot {\bf e}_{\k,\nu} (\bq) \bQ \cdot {\bf e}^{*}_{\k',\nu} (\bq) e^{i\bq \cdot {\bf R}_p}  \Big], \nonumber 
\end{eqnarray}
where  $M_\k$ and $M_0$ are the atomic and proton masses, and $\nu$ denotes the 
phonon branch index. The phonon polarization vectors, associated with phonon frequencies $\omega_{\bq \nu}$,
are denoted by ${\bf e}_{\k,\nu} (\bq)$ and the mode-resolved mean-square displacements of the atoms
are given by $u^2_{\bq \nu}  = \hbar/(2M_0 \omega_{\bq \nu})[2n_{\bq  \nu}(T) + 1 ]$, where $n_{\bq  \nu}(T)$ is 
the Bose-Einstein distribution. 
We emphasize that an important step in obtaining Eq.~\eqref{eqa1.7} exploits 
the translational symmetry of the lattice~\cite{Zacharias_arxiv_2021}.
Combining this point together with the partitioning of the phonons into 
two smaller Brillouin zone groups, $\mathcal{A}$ and $\mathcal{B}$~\cite{Giustino_2017},
allows for the efficient calculation of the total scattering intensity~\cite{Zacharias_arxiv_2021}. 
We also note that Eq.~\eqref{eqa1.7} and all subsequent expressions do not contain a constant prefactor that depends  
on the probe-sample interaction~\cite{Maradudin_Weiss,Grosso_Pastori_book}.

\begin{figure*}[htb!]
\includegraphics[width=0.90\textwidth]{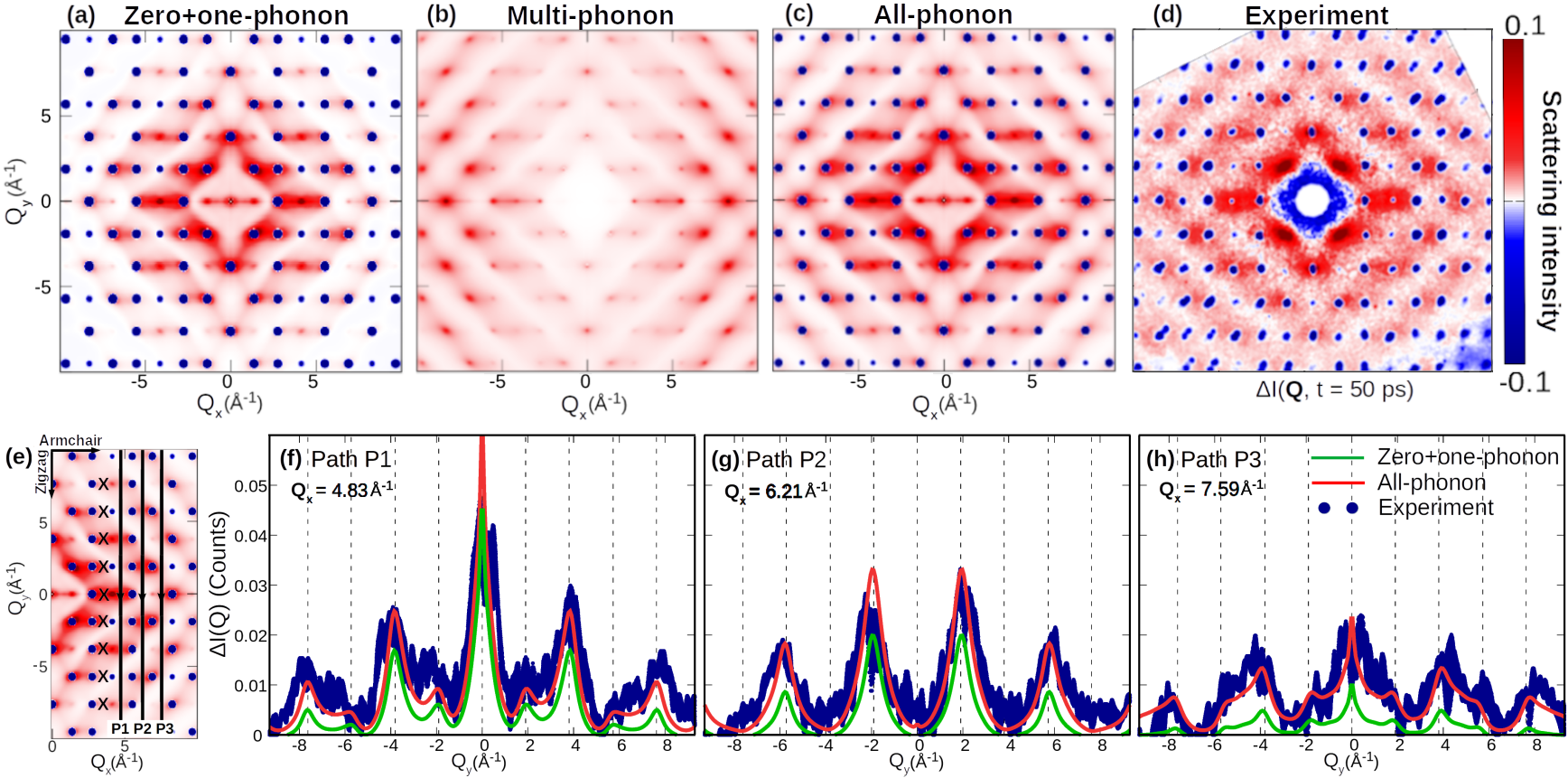}
\caption{
(a) Zero-plus-one-phonon ($\Delta I_0 + \Delta I_1$), (b) multi-phonon ($\Delta I_{\rm multi}$), and (c) all-phonon 
($\Delta I_{\rm all}$) difference scattering patterns of bulk bP calculated as 
$\Delta I (\bQ)  = I(\bQ,300 \,{\rm K}) - I(\bQ, 100\, {\rm K})$ to match the 
experimental conditions~\cite{Zahn_2020,Seiler2021}.
(d) Experimental difference scattering pattern of bulk bP measured at 50 ps from FEDS. 
Data is divided by the maximum count due to elastic scattering.
(e) All-phonon difference scattering pattern of bulk bP showing three vertical paths P1, P2, and P3  
at $Q_x=4.83$~\AA$^{-1}$, $6.21$~\AA$^{-1}$, and $7.59$~\AA$^{-1}$. Paths pass through several high-symmetry X-points 
in the zigzag direction. (f)-(h) $\Delta I (\bQ)$ as a function of $Q_y$ along P1, P2, and P3. 
Zero-plus-one-phonon, all-phonon, and experimental data is represented by green, red, and blue.
Vertical dashed lines indicate positions of high-symmetry X-points.
All calculated intensities were divided by the Bragg intensity at the zone centre, i.e $I_0(\bQ = {\bf 0})$, 
and multiplied by  the same scaling factor to facilitate comparison.  
The Brillouin zone was sampled using a $50\times50\times50$ $\bq$-grid and 
full patterns were obtained by a four-fold rotation. 
\label{fig2} 
}
\end{figure*}

Evaluation of the phononic factor, $e^{P_{p,\k\k'}}$, accounts
for all-phonon processes to the scattering intensity, including 
emission and absorption. Taking now the Taylor expansion
of $e^{P_{p,\k\k'}}$ in Eq.~\eqref{eqa1.7} and retaining the zeroth-order term 
we obtain the Bragg diffraction intensity as:
\begin{eqnarray}\label{eqa1.12_b}
I_0(\bQ,T) &=&   N_p^2 \sum_{\k \k'} f_\k (\bQ) f^*_{\k'} (\bQ)
    \text{cos}\big[ \bQ \cdot (\bt_\k - \bt_{\k'})\big] \nonumber  \\ 
&\times& e^{-W_{\k} (\bQ,T)} e^{-W_{\k'} (\bQ,T)}  \delta_{\bQ,\bG}, 
\end{eqnarray}
where we have employed the sum rule $\sum_{p} \exp(i\bQ \cdot {\bf R}_p) = N_p \, \delta_{\bQ,\bG}$. 
In a similar spirit, keeping the first order term yields the one-phonon scattering formula:
\begin{eqnarray}\label{eqa1.13}
&\,& I_1(\bQ,T) =  M_0 N_p \sum_{\k \k'} f_\k (\bQ) f^*_{\k'} (\bQ)
\frac{ e^{-W_{\k} (\bQ,T)}
   e^{-W_{\k'} (\bQ,T)}}{ \sqrt{M_\k M_{\k'}}} \nonumber  \\ &&
 \times   \sum_{\nu } \, \text{Re} \Big[ \bQ \cdot {\bf e}_{\k,\nu}(\bQ) 
   \bQ \cdot {\bf e}^{*}_{\k', \nu} (\bQ) e^{i\bQ \cdot [\bt_{\k'} - \bt_{\k}]} \Big] u^2_{\bQ  \nu}.
\end{eqnarray}
Subsequent higher-order terms in the expansion of $e^{P_{p,\k\k'}}$ correspond to inelastic excitations
of more than one phonon. Therefore, we write the all-phonon scattering intensity as a summation 
of the zero-, one- and multi-phonon terms, i.e.:  
\begin{eqnarray}\label{eq_all}
I_{\rm all}(\bQ,T) = I_0(\bQ,T) + I_1(\bQ,T) + I_{\rm multi}(\bQ,T)  . 
\end{eqnarray}

For our calculations we employed the unit cell of bP~\cite{Ribeiro_2018} with optimized lattice constants
$a =4.554$~\AA, $b= 3.307$~\AA, and $c =11.256$~\AA. 
The evaluation of the full set of phonon polarization vectors and frequencies was performed  
by means of density-functional perturbation theory (DFPT)~\cite{Baroni_2001} and 
Fourier interpolation as implemented in the {\tt Quantum ESPRESSO} suite~\cite{QE,QE_2}.
Using this information we calculate the Debye-Waller and phononic factors
to obtain $I_{\rm all}$, $I_{\rm 0}$, and $I_{\rm 1}$ from Eqs.~\eqref{eqa1.7}, ~\eqref{eqa1.12_b},
and~\eqref{eqa1.13}, respectively. The multi-phonon term is obtained as $I_{\rm multi} = I_{\rm all} - I_0 - I_1 $.
We remark that two-phonon, three-phonon, and subsequent contributions can be straightforwardly calculated 
by separating the appropriate order in the Taylor expansion of $e^{P_{p,\k\k'}}$ in Eq.~\eqref{eqa1.7}. 
The atomic scattering amplitude was evaluated as a sum of 
Gaussians~\cite{Vand_1957} using the parameters in Ref.~[\onlinecite{Peng_book}].
All patterns were calculated as the average of the scattering intensities 
in the $Q_x$-$Q_y$ planes at $Q_z = 0$ and $Q_z = 2 \pi / c = 0.56$~\AA$^{-1}$, where $Q_x$, $Q_y$, and $Q_z$ 
are the Cartesian components of $\bQ$.  
The code used for calculating the all-phonon scattering intensity and its various contributions 
is available at the {\tt EPW/ZG} tree~\cite{Ponce_2016_EPW,Zacharias_2020}. 
Full computational details are given in the parallel paper, Ref.~[\onlinecite{Zacharias_arxiv_2021}].

In Figs.~\ref{fig2}(a)-(d) we present our calculations of the difference 
scattering patterns of bP considering separate phonon contributions, 
and compare them with our measurements of the thermalized signals, 
all obtained as $\Delta I (\bQ)  = I(\bQ,300 \,{\rm K}) - I(\bQ, 100\, {\rm K})$. 
Our results indicate that thermal phonon populations exhibit a high degree of anisotropy, consistent with the
phonon band structure along the zigzag ($\Gamma$-A) and armchair ($\Gamma$-X) directions~\cite{2015Luo, Ribeiro_2018}. 
This finding reflects, essentially, the structural anisotropy of bP, giving rise to a different in-plane 
behavior of the thermal~\cite{2015Lee,2015Luo,2015Jang} and electrical conductivities~\cite{2014Qiao, 2014Xia,2014Liu,He2015}.
The calculated single-phonon scattering intensity [Fig.~\ref{fig2}(a)] is qualitatively in good agreement 
with experiment [Fig.~\ref{fig2}(d)] for $Q_x$ and $Q_y$ lying within $\pm 5$~\AA$^{-1}$. Beyond this range, 
the one-phonon map underestimates inelastic scattering missing clearly the outermost diamond-like features 
observed in the experiment. This discrepancy disappears once multi-phonon processes [Fig.~\ref{fig2}(b)]
are included as described in the LBJ theory. In fact, the calculated all-phonon scattering 
intensity [Fig.~\ref{fig2}(c)] reproduces the measured diffused pattern, suggesting that 
multi-phonon interactions dominate inelastic scattering processes with long wavevectors. 
Our analysis yields that the major contribution to 
multi-phonon scattering arises from two-phonon processes.
We stress that our calculations of the all-phonon LBJ scattering intensity have been verified in a straightforward 
fashion using Zacharias-Giustino (ZG) displacements~\cite{Zacharias_2016,Zacharias_2020}. As demonstrated in 
the parallel paper, Ref.~[\onlinecite{Zacharias_arxiv_2021}], although this method converges slower, ZG displacements provide the 
scatterers' coordinates that best reproduce Debye-Waller factors and all-phonon inelastic scattering.

\begin{figure}[t!]
\includegraphics[width=0.40\textwidth]{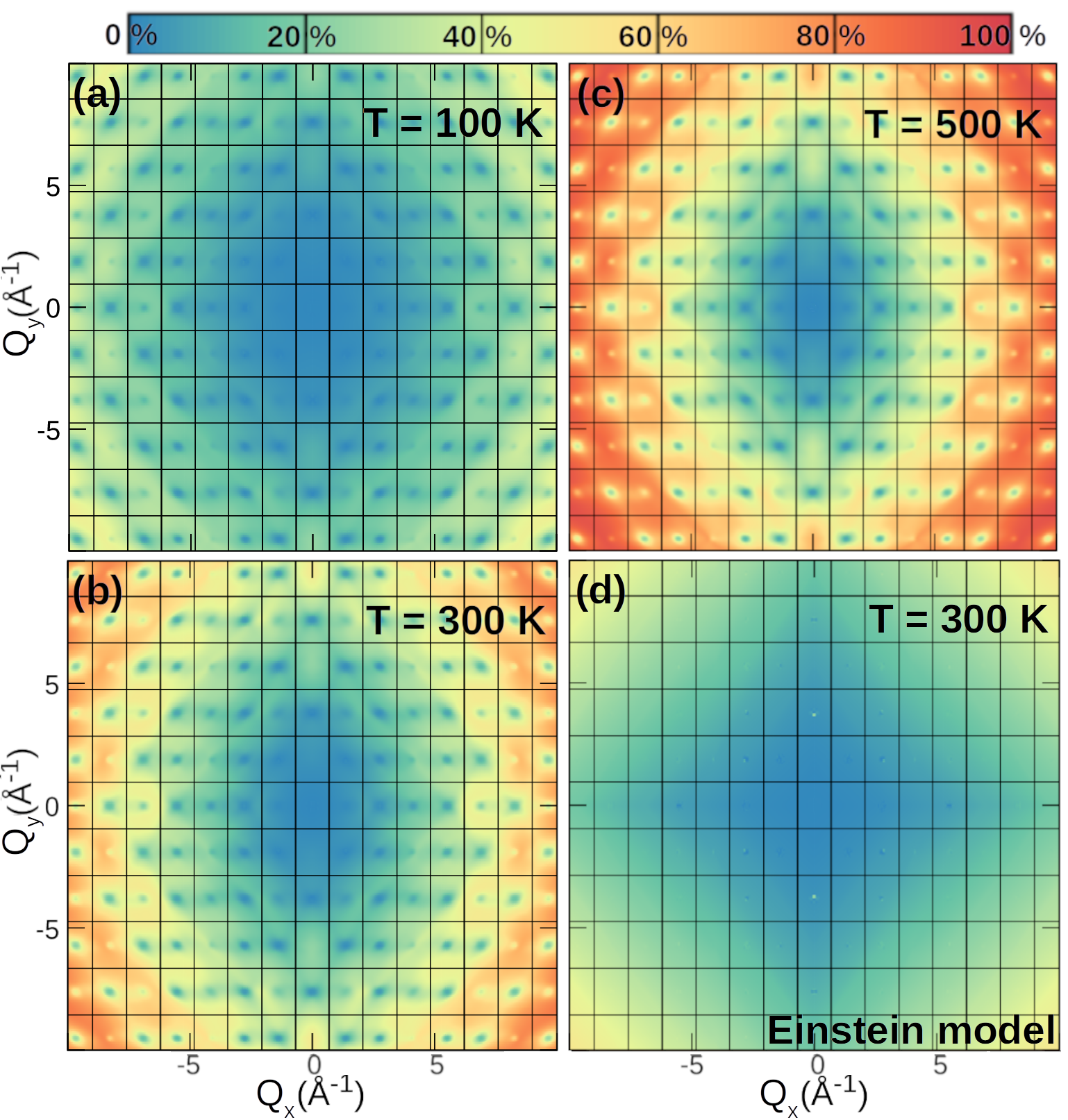}
\caption{
(a)-(c) Coloured maps showing the percentage contribution of multi-phonon interactions to thermal diffuse scattering, $\mathcal{P}$, 
across a wide range in reciprocal space of bP at 100~K, 300~K, and 500~K. (d) $\mathcal{P}_{\rm E}$ calculated within the
Einstein model for thermal diffuse scattering at 300~K.  Rectangles represent different Brillouin zones centered at $\Gamma+\bG$. 
\label{fig3} 
}
\end{figure}

Figures~\ref{fig2}(e)-(h) show a quantitative comparison between the theoretical and experimental thermal 
diffuse scattering intensities along three paths in the zigzag direction, passing through multiple high-symmetry X-points. 
These paths are labelled as P1, P2, and P3 in Fig.~\ref{fig2}(e) and are selected in a way
to (i) exclude elastic and thus focus on inelastic scattering, and (ii) explore
 inelastic scattering maxima in the experimental pattern. 
In Figs.~\ref{fig2}(f)-(h) we compare our calculated $\Delta I (\bQ, 300 \,{\rm K})$ with our measured
thermalized scattering intensity for paths P1, P2, and P3, respectively. Red and green curves represent the 
all-phonon and one-phonon scattering intensities; blue discs represent our measurements. 
As for Figs.~\ref{fig2}(a)-(d), all calculations are scaled by the same constant. 
The all-phonon scattering intensity is in excellent agreement with the experiment,
reproducing all main peak intensities at the X-points.  
The remaining discrepancies between the two sets of data can be attributed to multiple and Huang scattering, 
not included in our calculations, as well as the sample purity~\cite{Zhong_Lin_Wang,Sascha_2011,Zahn_2020}. 
It is also evident from Figs.~\ref{fig2}(f)-(h) that one-phonon processes are not 
sufficient to explain the all-phonon scattering intensity, especially for relatively large $|\bQ|$.
For example, along path P2 one-phonon processes contribute 60\% and 48\% to the main peaks at 
$Q_y = 1.9$~\AA$^{-1}$ and 5.7~\AA$^{-1}$. The intensity ratio
of these peaks is 54\% and 44\% for the all-phonon and one-phonon 
scattering, respectively. This comparison confirms that a single multiplicative factor is not adequate to 
explain the change in the scattering intensity calculated for the two cases. 
Furthermore, the one-phonon contribution becomes negligible for $Q_y > 7$~\AA$^{-1}$. 
A similar analysis can be applied for paths P1 and P3, demonstrating the significance of 
multi-phonon interactions in reproducing quantitatively and qualitatively the 
diffuse signals of bP over the full scattering vector range.

To clarify the role of multi-phonon interactions in bP we 
calculate the percentage $\mathcal{P} = I_{\rm multi}/ (I_{1} + I_{\rm multi})$ across
the full pattern for three different temperatures.
Figure~\ref{fig3} shows $\mathcal{P}$ as a function of $Q_x$ and $Q_y$ extending over 165 Brillouin zones
around the zone center calculated for (a) 100~K, (b) 300~K, and (c) 500~K. 
Our results reveal that the one-phonon theory can serve as a reliable method to analyze 
scattering signals from Brillouin zones that exhibit weak multi-phonon effects. However, 
even for $T = 100$~K, multi-phonon processes make a clear impact at large $|\bQ|$. As anticipated, 
$\mathcal{P}$ increases with temperature becoming more pronounced for regions closer to the center. 
It is also apparent that inelastic scattering around Bragg positions ($\Gamma+{\bf G}$ points) 
mostly originates from single-phonon interactions, even for large $|{\bf G}|$ and $T$. This observation 
is justified by considering that the low-frequency acoustic phonons dominate inelastic scattering 
at $\bQ \simeq \Gamma+{\bf G}$. To further quantify our results we evaluate the fraction of the 
vibrational energy transfer to the crystal due to multi-phonon excitations as: 
\begin{eqnarray}\label{eq_all}
\Delta \mathcal{E} (T) = \frac{\int_\bQ I_{\rm multi}(\bQ,T) d\bQ} {\int_\bQ I_1(\bQ,T) + I_{\rm multi}(\bQ,T)d\bQ},
\end{eqnarray}
 where the integrals are taken over the area of reciprocal space shown in Fig.~\ref{fig3}. 
We find $\Delta \mathcal{E}$ to be 10\%, 21\%, and 29\% at 100~K, 300~K, and 500~K, respectively. 

Now, we provide a metric that practically assesses the effect of multi-phonon interactions in any crystal. 
We employ the Einstein model and replace the phonon frequencies with their mean value $\omega_{\rm E}$, neglecting
dispersion, and set the associated eigenvectors to be isotropic~\cite{Hall_1965}.
Hence, the metric consists of evaluating the percentage $\mathcal{P}_{\rm E}$ and energy transfer $\Delta \mathcal{E}_{\rm E}$ 
using the Einstein model's analogues of $I_{1}$ and  $I_{\rm multi}$. These are obtained from the power series of 
$C_{\k\k'}(\bQ,T) = \bQ^2/( \omega_{\rm E}  \sqrt{M_\k M_{\k'}} )[2n_{{\rm E}}(T) + 1 ]$ using 
Eq.~(19) of Ref.~[\onlinecite{Zacharias_arxiv_2021}].
Keeping $|\bQ|$ and $T$ constants, the multi-phonon contribution to inelastic scattering depends on 
the material-specific values $M_\k$ and $ \omega_{\rm E}$. 
Figure~\ref{fig3}(d) shows $\mathcal{P}_{\rm E}$ calculated for bP at $T=300$~K using $\omega_{\rm E} = 279.7$ cm$^{-1}$.
At variance with the exact result in Fig.~\ref{fig3}(b), $\mathcal{P}_{\rm E}$ 
increases smoothly with $|\bQ|$ and lacks of any fine structure. Despite this shortcoming, 
our metric yields $\Delta \mathcal{E}_{\rm E} = 14$\% in good agreement with the actual value of $\Delta \mathcal{E}=21$\%.
In the parallel paper, Ref.~[\onlinecite{Zacharias_arxiv_2021}], we show for 2D MoS$_2$ that, although the mean phonon frequency 
($\omega_{\rm E} = 287.4$ cm$^{-1}$) is similar to that of bP, multi-phonon contributions are less pronounced giving 
$\Delta \mathcal{E}_{\rm E} = 10$\%. 
Based on our toy model, this difference is attributed to the large atomic mass of molybdenum being about three times 
larger than that of phosphorus. 

In conclusion, we have established a new first-principles method for the calculation of
the all-phonon inelastic scattering in solids based on the LBJ theory.  
The present work lays the foundations for developing a reverse engineering approach 
to extract nonequilibrium phonon populations from time-resolved experiments~\cite{Cotret_2019}. 
Identifying the all-phonon scattering signatures is also critical to apply sophisticated corrections on the 
experimental data and obtain reliable information of plasmon and magnetic excitations~\cite{Abajo_2010,Bowman_2019}. 
Our methodology can be upgraded to investigate polaron features~\cite{Adams_2000,Pengcheng_2000} 
and point defects~\cite{Krivoglaz1996} in diffused signals, study materials exhibiting anharmonic 
lattice dynamics~\cite{Hellman_2011,Errea_2014,Florian_2020}, as well as describe electron energy 
loss spectroscopy measurements~\cite{Lagos_2017,Hage_2018}. 
We stress that the adiabatic approximation employed here performs well
in most materials, however, in exceptional cases, such as 
highly-doped semiconductors, nonadiabatic effects cannot be ignored and
more sophisticated treatments, beyond standard DFPT, are required~\cite{Lazzeri_2006,Calandra_2010,Caruso_2017}. 
The present approach is suitable for both, condensed matter theorists and experimentalists, 
opening the way for systematic {\it ab-initio} calculations of phonon-induced inelastic scattering in solids.

Electronic structure calculations performed in this study
are available on the NOMAD repository~\cite{nomad_doi}.

\acknowledgments

M.Z. acknowledges financial support from the Research Unit of Nanostructured Materials Systems (RUNMS)
and the program META$\Delta$I$\Delta$AKT$\Omega$P of the Cyprus University of Technology.
H.S. was supported by the Swiss National Science Foundation under Grant No.~P2SKP2\textunderscore184100.
F.C. acknowledges funding from the Deutsche Forschungsgemeinschaft (DFG) - Projektnummer 443988403. 
F.G. was supported by the Computational Materials Sciences Program funded by the U.S. Department of Energy, 
Office of Science, Basic Energy Sciences, under Award DE-SC0020129.
R.E. acknowledges funding from the European Research Council (ERC) under the European Union’s Horizon 2020 
research and innovation program (Grant Agreement No. ERC-2015-CoG-682843) and by the Max Planck Society.
We acknowledge that the results of this research have been achieved
using the DECI resource Saniyer at UHeM based in Turkey~\cite{PRACE} 
with support from the PRACE aisbl and HPC resources from the Texas Advanced Computing Center 
(TACC) at The University of Texas at Austin~\cite{TACC}.

\bibliography{references}{}

\end{document}